\documentstyle[12pt]{article} 
\begin{document}
\begin{center}
{\bf Relativistic probability waves} \\[2cm] 
Marius Grigorescu \\[3cm]  
\end{center}
\noindent
$\underline{~~~~~~~~~~~~~~~~~~~~~~~~~~~~~~~~~~~~~~~~~~~~~~~~~~~~~~~~
~~~~~~~~~~~~~~~~~~~~~~~~~~~~~~~~~~~~~~~~~~~}$ \\[.3cm]
A canonical structure compatible with the action of the Lorentz group can be obtained considering the energy and time as conjugate variables of an extended phase space. Scalar probability waves, describing free relativistic particles, are associated with functional coherent states for an extended Liouville equation. Relativistic action waves are provided by distributions localized in the momentum space, evolving according to the continuity and Hamilton-Jacobi equations. Presuming the existence of minimum space and time intervals, the action distributions take the form of relativistic Wigner functions. The nonrelativistic quantum dynamics is retrieved approximating the time distribution function by a Gaussian wave packet. \\
$\underline{~~~~~~~~~~~~~~~~~~~~~~~~~~~~~~~~~~~~~~~~~~~~~~~~~~~~~~~~
~~~~~~~~~~~~~~~~~~~~~~~~~~~~~~~~~~~~~~~~~~~}$ 
\newpage
\section{ Introduction} 

The first evidence on the granular structure of the phase-space \cite{mpl} emerged from the study of the relativistic system represented by thermal radiation, within classical statistical mechanics. However, the usual Hamiltonian dynamics, of time-dependent coordinates and momenta, is not appropriate as a starting point for a Lorentz-covariant theory. \\ \indent
A canonical structure compatible with the action of the Lorentz group can be obtained extending the nonrelativistic phase space by energy and time, as conjugate variables \cite{macke}, setting
thereby the framework to relate electrodynamics, relativity, and quantum mechanics \cite{et}. These new variables have been considered also in a variational approach to the coupled Vlasov-Maxwell equations \cite{ajb}. In this work, relativistic probability waves are associated to the coherent solutions of the Liouville equation in the extended phase space. 
\\ \indent
The extended Hamiltonian dynamics of a classical particle is reviewed in Sect. 2. As the observable time becomes  a canonical coordinate, the evolution is described in terms of a parameter called universal time. The relativistic Liouville equation for the distribution function is presented in Sect. 3. Similarly to the nonrelativistic treatment \cite{cpw}, "action waves" are related to coherent functionals localized in the momentum space. For such probability distributions the Liouville equation reduces to the coupled continuity and Hamilton-Jacobi equations. It is shown that for a distribution localized in a finite domain of the extended phase-space, the universal time is the expectation value of the time coordinate in the "intrinsic frame". 
\\ \indent   
The transition from action distributions in the extended phase space, to Wigner functions, arising from a discontinuous character of the inertial motion, is discussed in Sect. 4. Following \cite{cpw}, discretization is introduced by presuming the existence of a minimum length $\ell$. Though, unlike the fundamental length $\ell_P$ at the Planck scale ($ \sim 10^{-35}$ m) used in string theory \cite{ven}, $\ell$ depends on the inertial parameter. The quantum "wave function" defined by discretization belongs to the extended Hilbert space presented in \cite{vpr}, and with respect to the universal time, it evolves according to a relativistic Schr\"odinger equation. In the "stationary" case this reduces to the Klein-Gordon equation, while the nonrelativistic limit is essentially "nonstationary". Within the present approach, the Wigner transform for relativistic quantum systems can be defined directly as a quasiprobability over the extended phase-space, rather than over trajectories \cite{pm}. Conclusions are summarized in Sect. 5.
\section{Classical dynamics in the extended phase space} 
The phase-space $M$ of a classical system can be extended to a phase-space $M^e$, which includes the energy and time as conjugate variables \cite{macke}. The canonical coordinates on $M^e$ consist of the canonical coordinates on $M$, denoted $\{q_i,p_i, i=1,n \}$, and $(q_0,p_0)$, supposed to be linear functions of time and energy, $q_0= c t$, respectively $p_0 = - E /c$, where $c$ is a dimensional constant, identified with the speed of light in vacuum \cite{et}. \\ \indent
Let $u$ be the "universal time" parameter along the trajectories on $M^e$, $d_u \equiv d /d u $ the derivative with respect to $u$, and
\begin{equation}
{\cal L}_{H^e} = \sum_{i=1}^n (d_u q_i) \partial_{q i} + (d_u p_i) \partial_{p i} + (d_u t ) \partial_t + (d_u E) \partial_E \label{ld}
\end{equation}
the Lie derivative ${\cal L}_{H^e} f \equiv - \{ H^e , f \}^e $. Here $\{ * , * \}^e$ and $H^e$ are the extended Poisson bracket and Hamilton function, respectively, on $M^e$. In the case of a nonrelativistic system $H^e$ can be taken of the form 
\begin{equation}
H^e_N= H+cp_0~~,
\end{equation}
where $H$ is the usual Hamilton function defined on $M$. For this expression, the corresponding equations of motion in the extended phase-space are
\begin{equation} 
d_u q_i = \frac{ \partial H}{ \partial p_i} ~~,~~ d_u p_i = - \frac{ \partial H}{ \partial q_i} 
\end{equation} 
\begin{equation} 
d_u t = 1 ~~,~~ d_u E = \frac{ \partial H}{ \partial t } ~~.
\end{equation} 
The first group of equations reproduces the usual Hamilton equations on $M$. The second group shows that the choice of $H^e_N$ corresponds to $u =t$, and ensures the conservation of the energy when $H$ is independent of time. \\ \indent
In the extended phase-space, the transition from Newtonian to relativistic mechanics (recalled in Appendix 1) consists essentially in a change of Hamiltonian.  
A free relativistic particle\footnote{To include an external potential ${\rm V}({\bf q})$ we may consider $H^e = - c \sqrt{(p_0+{\rm V}/c)^2 - {\bf p}^2}$.} having $q^e \equiv (q_0,{\bf q})$ and  $p^e \equiv (p_0,{\bf p})$ as phase-space coordinates on $M^e \equiv {\sf R}^8$, can be described by $H^e_0= - c \sqrt{p_0^2 - {\bf p}^2}$. When $p_0^2 \approx m_0^2c^2 \gg {\bf p}^2$ this extended Hamiltonian reduces to the nonrelativistic expression $H^e_{N0}={\bf p}^2/2m_0 +c p_0$, while in general it provides the equations of motion 
\begin{equation} 
d_u q_0 = - c \frac{p_0}{ \sqrt{p_0^2 - {\bf p}^2}}  ~~,~~ d_u p_0 = 0  \label{cle0}
\end{equation} 
\begin{equation} 
d_u {\bf q} = c \frac{ {\bf p}}{ \sqrt{p_0^2 - {\bf p}^2}} ~~,~~ d_u {\bf p} =0 ~~.
\label{cle1}
\end{equation} 
With respect to a particular inertial frame, the usual velocity\footnote{The usual Lagrangian on ${\sf R}^3$ of the relativistic free particle can be found in \cite{am}.} ${\bf v}= d {\bf q} / d t $ is the ratio ${\bf v}= c d_u {\bf q} / d_u q_0 = - c {\bf p} / p_0$.  \\ \indent
The inertial parameters ${I_\mu}$ for $H^e_0$, defined by 
\begin{equation}
\frac{1}{I_\mu} \equiv \frac{1}{p_\mu} \frac{\partial H^e_0}{\partial p_\mu}~~,~~\mu=0,1,2,3
\end{equation}
take the values
\begin{equation}
I_1 =I_2=I_3 = -I_0 = - \frac{H^e_0}{c^2} = m_0~~, 
\label{in}
\end{equation}
as provided by the invariant value $-m_0c^2$ of $H^e_0$.   
 
\section{ Relativistic action waves} 

The statistical properties of classical systems composed of $N$ identical relativistic particles can be described by a distribution function $f^e \ge 0$, depending on $u$, defined on the one-particle extended phase-space $M^e$. If $d \Omega_{m^e}$ denotes the volume element around the point $m^e \in M^e$, then $f^e(m^e,u)$ is normalized using the integrality condition   
\begin{equation}
\int d \Omega_{m^e} f^e(m^e,u) = N~~,~~N \geq 1~~. \label{nc}
\end{equation}
For macroscopic systems, the probability to find a particle localized in $d \Omega_{m^e}$,
proportional to $f^e(m^e,u) \delta \Omega_{m^e}$, is given essentially by the particle density in $d \Omega_{m^e}$, while at small $N$ we can expect a definition in terms of the average universal time interval of localization in $d \Omega_{m^e}$. \\ \indent
Let us consider a system consisting of a single relativistic particle ($N=1$), with $d \Omega_{m^e} \equiv d^4q d^4p$, ($d^4p \equiv dp_0 d^3p$, $d^4 q \equiv dq_0 d^3 q$), and Hamiltonian $H^e_0(p^e)= - c \sqrt{p_0^2 - {\bf p}^2}$. The distribution function $f^e(q^e,p^e,u)$ evolves according to the relativistic Liouville equation
\begin{equation}
\partial_u f^e + {\cal L}_{H^e_0} f^e =0 \label{lieq}
\end{equation}
where ${\cal L}_{H^e_0}$ provided by (\ref{ld}-\ref{cle1}) has the form
\begin{equation}
{\cal L}_{H^e_0} =  \frac{ c {\bf p} \cdot {\bf \nabla}}{ \sqrt{p_0^2 - {\bf p}^2 }}
- \frac{c p_0 \partial_0}{ \sqrt{p_0^2 - {\bf p}^2}}  ~~,
\end{equation}
with $\partial_0 \equiv \partial / \partial q_0$ and $\nabla \equiv \partial / \partial {\bf q}$. Thus, (\ref{lieq}) becomes
\begin{equation}
\sqrt{p_0^2 - {\bf p}^2} \partial_u f^e - c p_0 \partial_0 f^e + c {\bf p} \cdot {\bf \nabla} f^e =0 ~~. \label{leq0}
\end{equation}
To find coherent solutions of this equation it is convenient to use the Fourier transform  
\begin{equation}
\tilde{f}^e(q^e,k^e, u) \equiv \int d^4 p  ~e^{i k_0p_0 + i{\bf k} \cdot {\bf p}} f^e(q^e, p^e, u) ~~, \label{fk}
\end{equation}
which provides the spectrum of the "momentum frequencies", $k^e/ 2 \pi$, $k^e \equiv (k_0,{\bf k})$. If $\sqrt{p_0^2 - {\bf p}^2}$ can be expressed in the form $m_0 c + \delta m_0 c$, where $\delta m_0$ as a function of $p_0^2$ and ${\bf p}^2$ is a power series, then by Fourier transform (\ref{leq0}) becomes
\begin{equation}
\tilde{H}^e_0 \partial_u \tilde{f}^e -i c^2 \partial_{k_0} \partial_0 \tilde{f}^e + i  c^2 \nabla_{\bf k} \cdot \nabla \tilde{f}^e =0~~, \label{fle}
\end{equation}
where $\partial_{k 0} \equiv \partial / \partial k_0$, $\nabla_{\bf k} \equiv \partial / \partial {\bf k}$, and  formally
\begin{equation}
\tilde{H}^e_0 = -c \sqrt{ (-i \partial_{k_0})^2-  (-i \nabla_{\bf k})^2 } ~~. \label{he}
\end{equation}
Various densities in space-time, such as the localization probability $n^e$, or current $J_\mu$, can be expressed directly in terms of $\tilde{f}$ and its partial derivatives $\partial_{k \mu} \equiv \partial / \partial k_\mu$, $\mu =0,1,2,3$, at $k^e= 0$ by
\begin{equation}
n^e(q^e,u) \equiv \int d^4p ~ f^e(q^e, p^e, u)  = \tilde{f}^e (q^e,0, u)
~~,  \label{n}
\end{equation}
\begin{equation}
J_\mu (q^e,u) \equiv \frac{1}{m_0} \int d^4p ~ p_\mu f^e(q^e, p^e, u)  = - \frac{i}{m_0} \partial_{k \mu} \tilde{f}^e (q^e,0, u)~~. \label{av1}
\end{equation}
In general, the mean value of an observable ${\cal O}(q^e, p^e)$, which is a polynomial as a function of the momentum components, has the expression
\begin{equation}
\langle {\cal O} \rangle (u) \equiv \int d^4q d^4p ~ {\cal O} f^e(u) = \int d^4q ~ {\cal O}(q^e, - i \partial_{k^e})  \tilde{f}^e(q^e, 0, u)~~.     
\end{equation} \indent
A particular class of coherent solutions for the relativistic Liouville equation (\ref{leq0}) consists of the "action distributions"  
\begin{equation}
f_0^e(q^e,p^e,u) = n^e(q^e,u) \delta( p_0 - \partial_0 S ) \delta( {\bf p} - \nabla S ) ~~, \label{cs1}
\end{equation}
where $n^e$ is the localization probability density in space-time, and $S(q^e,u)$ is the generating function of the Hamilton-Jacobi theory. By Fourier transform (\ref{cs1}) becomes 
\begin{equation}
\tilde{f}_0^e = n^e e^{i k_0 \partial_0 S+i {\bf k} \cdot \nabla S }   \label{f0}
\end{equation}
while (\ref{fle}) reduces to the system of equations 
\begin{equation}
\partial_u [ n^e \sqrt{ (\partial_0 S)^2 - (\nabla S)^2}~ ] =c \partial_0 (n^e \partial_0 S) - c \nabla \cdot (n^e \nabla S) \label{cont} 
\end{equation}
and 
\begin{equation}
n^e \partial_\mu {\cal J} = 0~~,~~ \mu=0,1,2,3 
\end{equation}
where $\partial_\mu \equiv \partial / \partial q_\mu$, ${\cal J} = \partial_u S - c \sqrt{(\partial_0 S)^2 - (\nabla S)^2} $. Thus, the solution ${\cal J} =0$ is nothig but the Hamilton-Jacobi equation in the extended phase-space,
\begin{equation}
\partial_u S + H^e_0 ( \partial_0 S, \nabla S)=0 \label{hj}~~.
\end{equation}
Considering $\partial_u S = m_0c^2$, (\ref{hj}) takes the form  
\begin{equation}
(\partial_0 S)^2 - (\nabla S)^2 = m_0^2 c^2~~,
\end{equation}
and (\ref{cont}) becomes the continuity equation
\begin{equation}
m_0 \partial_u n^e = \partial_0 (n^e \partial_0 S) -  \nabla \cdot (n^e \nabla S)~~. \label{cont1}
\end{equation}
In terms of the density (\ref{n}), the mean value of the time coordinate is
\begin{equation}
\langle t \rangle = \frac{1}{c} \int d^4q~ q_0~ n^e ~~,
\end{equation}
so that
\begin{equation}
d_u \langle t \rangle  = \frac{1}{c} \int d^4q~ q_0~ \partial_u n^e ~~. \label{dut}
\end{equation}
Let us presume that $n^e$ is limited in time, confined to a finite volume ${\cal V}$ in space, and $\nabla S$ vanishes along the normal to the boundary of ${\cal V}$. In this case (\ref{cont1},\ref{dut}) yield 
\begin{equation}
d_u \langle t \rangle  =- \frac{1}{m_0 c} \int dq_0 \int_{\cal V} d^3q~ n^e \partial_0 S = \frac{\langle E \rangle }{m_0c^2} \label{dt} 
\end{equation}
where $\langle E \rangle $ is the mean value of the energy. Because $\langle E \rangle $ is a positive constant\footnote{As required to ensure the Lorentz action (\ref{lga}) \cite{et}.}, (\ref{dt}) shows that $\langle t \rangle $ and $u$ are in the linear relationship 
\begin{equation}
\langle t \rangle = \frac{\langle E \rangle}{m_0c^2} u~~. \label{tu}
\end{equation}
The localization to a finite domain in space-time makes possible to define (up to a translation) an "intrinsic frame" (IF), as the frame selected by the condition $\langle {\bf p} \rangle_{\rm IF} =0$. Expressed in terms of the intrinsic expectation values, (\ref{tu}) shows that the universal time $u$ corresponds up to the factor $m_0 c^2 / \langle E \rangle_{\rm IF} $ to the mean time in the intrinsic frame, $\langle t \rangle_{\rm IF}$. Thus, if we can define $m_0$ as $\langle E \rangle_{\rm IF}/c^2$, then $u= \langle t \rangle_{\rm IF}$. For a density   $n^e(q^e,u) = \delta (q_0-c u) n({\bf q},u)$, localized in time, (\ref{cont1}) reduces in the nonrelativistic limit to the usual continuity equation $\delta (t -u) [m_0 \partial_u n +  \nabla \cdot (n \nabla S)] =0$. 
\section{The relativistic Schr\"odinger equation} 
When the partial derivatives $\partial_\mu S$, $\mu=0,1,2,3$, in (\ref{f0})
are written as finite differences $[S(q^\mu + \ell_\mu/2,u)- S(q^\mu - \ell_\mu/2,u)]/ \ell_\mu$, the action distribution becomes    
\begin{equation}
\tilde{f}_0^e(q^e,k^e,u) = \lim_{\sigma_\mu \rightarrow 0} 
\Psi (q^\mu + \frac{\sigma_\mu k_\mu}{2},u) \Psi^*(q^\mu - \frac{\sigma_\mu k_\mu}{2},u) \label{lsg}~~,
\end{equation}
where $\sigma_\mu$ denotes the ratio $\sigma_\mu \equiv \ell_\mu / k_\mu$, and
$\Psi(q^e,u)$ is the complex function $\Psi = \sqrt{n^e} \exp( i S / \sigma_\mu )$.
However, if the limit of $\sigma_\mu$ when both $\ell_\mu$ and $k_\mu$ decrease to zero is finite, having the same value $\sigma$ for all components, then
we may consider also "quantum distributions" 
\begin{equation}
\tilde{f}^e_\Psi (q^e,k^e,u)  \equiv \Psi (q^\mu + \frac{\sigma k^\mu}{2},u) \Psi^*(q^\mu - \frac{\sigma k^\mu}{2},u) ~~,  
\label{fq}
\end{equation} 
as possible functional coherent states for (\ref{fle}). In this case, the normalization condition (\ref{nc}) for the corresponding phase-space distribution   $f^e_\Psi$ takes the form 
\begin{equation}
\int d^4q d^4p f^e_\Psi (q^e,p^e,u) = \int d^4q  \vert \Psi (q^e,u) \vert^2 =1 
\end{equation} 
and the phase-space overlap between two distributions $f^e_{\Psi_1}$, $f^e_{\Psi_2}$ is 
\begin{equation}
<f^e_{\Psi_1} f^e_{\Psi_2} > \equiv \int d^4q d^4p f^e_{\Psi_1} f^e_{\Psi_2} = \frac{ \vert \langle \Psi_1 \vert \Psi_2 \rangle \vert^2}{(2 \pi \sigma)^4} 
\end{equation} 
where 
\begin{equation}
\langle \Psi_1(u_1) \vert \Psi_2(u_2) \rangle \equiv \int d^4q \Psi_1^*(q^e,u_1) \Psi_2(q^e,u_2)  ~~. \label{ampl}
\end{equation} \indent 
A linear relationship $\ell_\mu = \sigma k_\mu$ with a finite, isotropic, Lorentz-invariant, universal phase-space element $\sigma$ could be related in principle to the existence of minimum space and time intervals for a certain energy domain, but is more difficult to justify than in the nonrelativistic case \cite{cpw}. In electrodynamics we can find limits such as the classical electron radius $r_e= \alpha \hbar / m_e c = 2.8$ fm, and the related cutoff energy $e^2/4 \pi \epsilon_0 r_e = m_e c^2$. In general, we can note that in the intrinsic frame, after a cutoff at $\approx 3 m_0c^2$ of the energy range (Appendix 2), $f^e_\Psi$ still remains unchanged in 94\% of the velocity domain $[0,c)$. With this approximation, the inverse of (\ref{fk}) can be replaced by a multiple Fourier series in which the components of $k^e$ from the factor $\exp(- i k_0p_0 - i{\bf k} \cdot {\bf p})$ take an infinite set of discrete values separated by $\kappa = \pi/ 3m_0c$. Also, if the time distribution has the variance $\delta t^2_0$, then the  intervals of ordered, physical time (e.g. the lifetime $\tau_L = \hbar / \Gamma$ for unstable particles) are greater than $\delta t_0$, and the length between any two fixed endpoints along a trajectory parametrized by $\langle t \rangle$, greater than $\ell = c \delta t_0$. Thus, a finite value $\sigma = \ell / \kappa \sim  m_0c^2 \delta t_0 $ should be expected. For a quantum particle, with $\sigma = \hbar$ we get $\delta t_0 \sim \hbar /m_0c^2$ and $\ell \sim \hbar /m_0c$, proportional to the inverse of the mass. \\ \indent
It is interesting to remark that beside the formal arguments, evidence for the physical relevance of the interval $\hbar /m_0c^2$ arises from the particle data. The values obtained for the ratio $ m_0c^2 / \Gamma \sim \tau_L / \delta t_0$ between the mass (in MeV) and decay width ($\Gamma$), using the experimental data \cite{rpp} for meson and baryon resonances are represented in Figure 1, (A) and (B), respectively. These values are well interpolated by functions of the form $2.1 + C /\Gamma$, where $C$ is 1222 MeV for mesons and  1487 MeV for baryons. Thus, $\tau_L$ appears to be limited below by $2 \hbar /m_0c^2$. \\ \indent
Denoting $\tilde{f}^e_\Psi \equiv (  \hat{U} \Psi)  (\hat{U}^{-1} \Psi^*)  $, 
with  $\hat{U} = \exp [ \sigma ( k_0 \partial_0+ {\bf k} \cdot \nabla)/2]$, 
 (\ref{fle}) becomes
\begin{equation}
\tilde{H}^e_0 \partial_u \tilde{f}^e_\Psi  = i \frac{\sigma c^2}{2}[ (  \hat{U} \Box \Psi)  (\hat{U}^{-1}  \Psi^*) -(  \hat{U} \Psi)  (\hat{U}^{-1} \Box \Psi^*) ]~~, \label{dufpsi}
\end{equation}
where $\Box \equiv \partial_0^2 - \nabla^2$. \\ \indent 
A "static" distribution $\partial_u \tilde{f}^e_\Psi =0$ is obtained if  $\Box \Psi = a \Psi$, where $a$ is a real constant. This constant can be estimated by using the expectation values of $H^e_0$ or $(H^e_0)^2 $. For simplicity, $\langle (H^e_0)^2 \rangle = m_0^2c^4 $ means
\begin{equation}
\int d^4q d^4p ~(p_0^2-{\bf p}^2 - m_0^2c^2) f^e_\Psi(q^e,p^e,u) =0 \label{inv}
\end{equation}
or 
\begin{equation}
\int d^4q [ (\hat{\cal K} \Psi) \Psi^* + \Psi (\hat{\cal K} \Psi^*)] =0~~, \label{const}
\end{equation}   
where
$\hat{\cal K} = - \sigma^2 \Box -m_0^2c^2$. When $\Box \Psi = a \Psi$, (\ref{const}) yields $ a \sigma^2 =- m_0^2c^2$, so that $\hat{\cal K} \Psi =0$, or
\begin{equation}
- \sigma^2 \Box \Psi = m_0^2c^2 \Psi~~. \label{kg}
\end{equation}
In the quantum theory this represents the Klein-Gordon equation \cite{bd}. Although all particles described by (\ref{kg}) are unstable, closer to stability are quark-antiquark systems like the $\pi^\pm$ and $K^\pm$ mesons\footnote{The singlet (triplet) states of $e^-e^+$ positronium have a lifetime of 1.2 ns (140 ns).} with a lifetime $\sim 10$ ns, much larger than $\hbar/m_0c^2  \sim 10^{-24}$ s.  
\\ \indent
In the nonstationary case (\ref{dufpsi}) becomes
\begin{equation}
\tilde{H}^e_0 [ (  \hat{U} i \partial_u \Psi) ( \hat{U}^{-1} \Psi^*) -(  \hat{U} \Psi)  \hat{U}^{-1} (i \partial_u \Psi)^*)] = 
\end{equation}
$$
- \frac{\sigma c^2}{2}[ (  \hat{U} \Box \Psi)  (\hat{U}^{-1}  \Psi^*) -(  \hat{U} \Psi)  (\hat{U}^{-1} \Box \Psi^*) ]~~, 
$$
or
\begin{equation}
 \tilde{H}^e_0  \Psi^*_-  i \partial_u \Psi_+  = 
- \frac{\sigma c^2}{2} \Psi^*_-   \Box \Psi_+   ~~, \label{rse} 
\end{equation}
\begin{equation}
 \tilde{H}^e_0  \Psi_+  i \partial_u \Psi_-^*  = 
\frac{\sigma c^2}{2} \Psi_+  \Box \Psi^*_-   ~~, \label{rse-} 
\end{equation}
where $\Psi_+ \equiv \hat{U}  \Psi $ and $\Psi^*_- \equiv \hat{U}^{-1}  \Psi^* $.
By contrast to the nonrelativistic case \cite{cpw}, in general (\ref{rse}-\ref{rse-}) cannot be reduced to separate equations for $\Psi_+ $ and $\Psi^*_-$ due to $\tilde{H}^e_0$, which acts on both functions. However, $\Psi_+ $ and $\Psi^*_-$ become complex conjugate at $k^e=0$, so that when $\sigma = \hbar$, the limit 
\begin{equation}
\lim_{k^e \rightarrow 0}  (\Psi^*_-)^{-1}  \tilde{H}^e_0  \Psi^*_-  \hat{U} i \partial_u \Psi  =  - \frac{\hbar c^2}{2}  \Box \Psi   ~~, \label{rse1} 
\end{equation}
can be formally considered as a relativistic Schr\"odinger equation for the wave function $\Psi$. Nonstationary solutions of this equation correspond for instance to wave-packets  of the form 
\begin{equation}
\Psi(q^\mu,u)= \chi (q_0,u) \psi({\bf q},u) \label{wf} 
\end{equation}
where $\chi (q_0,u) \equiv \chi_{{\cal Q}_0,{\cal P}_0} (q_0,u) $ is a Glauber coherent state \cite{vpr}
\begin{equation}
\chi (q_0,u) = \sqrt{ \frac{ \Omega}{c \sqrt{ \pi}}}  e^{- \Omega^2 (q_0- {\cal Q}_0)^2/2c^2 + i {\cal P}_0  (q_0- {\cal Q}_0 /2)/ \sigma }~~, \label{chi}
\end{equation}
with the centroid at ${\cal Q}_0 = - u {\cal P}_0 / m_0 $ and variance $c^2/2 \Omega^2$. The parameters ${\cal P}_0,{\cal Q}_0$ are related to the energy and time
expectation values $\langle E \rangle,  \langle t \rangle$ by the relations  ${\cal P}_0 \equiv \langle p_0 \rangle =- \langle E \rangle /c$, respectively ${\cal Q}_0 \equiv \langle q_0 \rangle  =c \langle t \rangle$.  \\ \indent
The function $f^e_\Psi(q^e, p^e, u)$ defined by $\tilde{f}^e_\Psi (q^e,k^e,u)$ inveting (\ref{fk})  is\footnote{Because $f^e_\Psi$ remains finite for $\vert p_0 \vert < m_0c$, (\ref{chi}) should be regarded as an approximation.}
\begin{equation}
f^e_\Psi(q^e, p^e, u) = \frac{1}{\pi \sigma}  e^{- \Omega^2 (q_0- {\cal Q}_0)^2/c^2 
- c^2 (p_0- {\cal P}_0)^2/ \Omega^2 \sigma^2} f_\psi({\bf q},{\bf  p}, u)
\end{equation}
where $f_\psi({\bf q},{\bf  p}, u)$ is the usual Wigner transform of $\psi( {\bf q},u)$. Thus, the time variance $\delta t^2 \equiv \langle t^2 \rangle- \langle t \rangle^2  = 1/ 2 \Omega^2$, as well as the energy variance $\delta E^2 = c^2 \delta p_0^2$,   $\delta p_0^2 \equiv  \langle p_0^2 \rangle- \langle p_0 \rangle^2  = \sigma^2 \Omega^2 /2 c^2$, are both finite, and satisfy the uncertainty relation $\delta E \delta t = \sigma /2$. 
 \\ \indent
With (\ref{chi}), the dependence on $k_0$ in $\tilde{f}^e_\Psi (q^e,k^e,u)$ can be separated in 
\begin{equation}
\tilde{g}_0(k_0) = e^{- \delta p_0^2 k_0^2 /2 +i {\cal P}_0 k_0 }~~, 
\end{equation}
so that 
\begin{equation}
\tilde{f}^e_\Psi (q^e,k^e,u)= \tilde{g}_0(k_0) \vert \chi (q_0,u) \vert^2 
(\hat{U}_{\bf k} \psi)  (\hat{U}^{-1}_{\bf k} \psi^*)~~,
\end{equation}
and (\ref{rse}) becomes
\begin{equation}
\tilde{H}^e_0   {\cal F}_k \chi^*  \hat{U}_{\bf k} i \partial_u \Psi  = 
- \frac{\sigma c^2}{2}  {\cal F}_k \chi^* \hat{U}_{\bf k} \Box \Psi   ~~. \label{eu} 
\end{equation}
Here $\hat{U}_{\bf k} = e^{\sigma {\bf k} \cdot \nabla /2}$ and ${\cal F}_k$ denotes the function ${\cal F}_k ({\bf q},u) =\tilde{g}_0(k_0)\hat{U}^{-1}_{\bf k}  \psi^*({\bf q},u) $. \\ \indent
For nonrelativistic particles we can expect that $\psi$ evolves over a time-scale much larger than $1/\Omega$, and approximate solutions of (\ref{eu}) can be obtained by taking the average over $q_0$. Using the equalities $\partial_u \Psi = (\partial_u \chi) \psi + \chi (\partial_u \psi) $,
$$
\int dq_0 \chi^* (q_0,u) \partial_u \chi (q_0,u) = - \frac{i {\cal P}_0}{2 \sigma} d_u{\cal Q}_0 =  \frac{{i \cal P}_0^2 }{2  \sigma m_0} ~~,
$$
and
\begin{equation}
\langle p_0^2 \rangle = \int dq_0 \chi^* (q_0,u) (- \sigma^2 \partial_0^2) \chi (q_0,u) = {\cal P}_0^2+ \delta p_0^2~~, \label{p02} 
\end{equation}
the integration over $q_0$ in both sides of (\ref{eu}) yields
\begin{equation}
\tilde{H}^e_0 {\cal F}_k  \hat{U}_{\bf k}( i \sigma \partial_u \psi
- \frac{{\cal P}_0^2}{2m_0} \psi)  = \frac{ c^2}{2} {\cal F}_k \hat{U}_{\bf k} (\langle p_0^2 \rangle + \sigma^2 \nabla^2 ) \psi  ~~. \label{eu1} 
\end{equation}
In general, $\tilde{H}^e_0 {\cal F}_k  \hat{U}_{\bf k}$ is a complicated operator because $\tilde{H}^e_0$ of (\ref{he}) introduces mixed partial derivatives, acting both on $\tilde{g}_0(k_0)$ and $\hat{U}_{\bf k}$. However, in the limit $k^e \rightarrow 0$  
$$
\tilde{H}^e_0 ~\tilde{g}_0(k_0) \approx -c \sqrt{ \langle p_0^2 \rangle+ \nabla_{\bf k}^2 }~ \tilde{g}_0(k_0) 
$$
where, according to the constraint (\ref{inv}), $\langle p_0^2 \rangle = m_0^2c^2 + \langle {\bf p}^2 \rangle$. Moreover, because 
$$
\lim_{k^e \rightarrow 0} \int d^3 q ( \langle {\bf p}^2 \rangle +  \nabla_k^2) {\cal F}_k  \hat{U}_{\bf k} \psi 
= \langle {\bf p}^2 \rangle  + \sigma^2 \int d^3 q ~ \psi^* \nabla^2  \psi =0 ~~,
$$ 
we approximate $\tilde{H}^e_0 {\cal F}_k  \hat{U}_{\bf k} \approx - m_0c^2 {\cal F}_k \hat{U}_{\bf k}$, so that when $k^e \rightarrow 0$  (\ref{eu1}) becomes \begin{equation}
i \sigma \partial_u \psi- \frac{{\cal P}_0^2}{2m_0} \psi  =- \frac{1 }{2m_0} (\langle p_0^2 \rangle + \sigma^2 \nabla^2 ) \psi  ~~. \label{eu2} 
\end{equation}
Using (\ref{p02}), this reduces further to
\begin{equation}
i \sigma \partial_u \psi  = - \frac{1}{2m_0} ( \delta p_0^2 + \sigma^2 \nabla^2 ) \psi  ~~.
\end{equation}
The term $ \delta p_0^2 /2 m_0 \equiv  (\sigma \Omega)^2/4m_0c^2$ can be included in a global $u$-dependent phase-factor of $\psi$,  while by changing the parametrization to the  mean time $\langle t \rangle = {\cal Q}_0 /c $,  one obtains
\begin{equation}
i \sigma \partial_{\langle t \rangle} \psi  = \frac{c}{2 {\cal P}_0} \sigma^2 \nabla^2  \psi  ~~.
\end{equation}
Here, according to (\ref{p02}), ${\cal P}_0 = - \sqrt{m_x^2 c^2 + \langle  {\bf p}^2 \rangle } $ with $m_x = \sqrt{m_0^2 - \delta p_0^2 /c^2}$ the effective mass, so that 
\begin{equation}
i \sigma \partial_{\langle t \rangle} \psi  =  - \frac{ \sigma^2  \nabla^2 }{2m_x}  ( 1 - \frac{ \langle  {\bf p}^2 \rangle }{2 m_x^2 c^2})  \psi   ~~.
\end{equation} \indent
In the case of an atomic electron ($\sigma = \hbar$, $m_x=m_e$, 1 a.u. $=\alpha^2 m_e c^2$), the correction term $\langle H_c \rangle=\hbar^2   \langle  {\bf p}^2 \rangle \langle \nabla^2 \rangle / 4 m_e^3 c^2$ can be compared with the usual contribution due to the variation of mass with velocity, $\langle H'_1 \rangle =-  \langle p^4 \rangle /$ $ 8 m_e^3 c^2$ \cite{ame}. For the ground state of hydrogen, $\langle H_c \rangle  =  - \alpha^2 /4$ a.u., while $\langle H'_1 \rangle =- 5 \alpha^2/8$ a.u.. The whole correction to this order found by expanding in powers of $\alpha$ the exact solution of the Dirac equation is  $- \alpha^2/8$ a.u.  \cite{ame}. 
\\ \indent
The interval $2 \delta t = \sqrt{2} / \Omega \equiv \hbar / \delta E$, is a measure of the time shift $\vert {\cal Q}_0'-{\cal Q}_0 \vert/c$ for which the overlap $\vert \langle \chi_{{\cal Q}_0',{\cal P}_0} \vert \chi _{{\cal Q}_0,{\cal P}_0} \rangle \vert^2$ between two states (\ref{chi}), and the corresponding transition amplitude (\ref{ampl}), remain significant. In general, this is much larger than $\delta t_0  \sim \hbar /m_0c^2$. For instance, if we take $\delta E \approx \epsilon_r m_xc^2$, where $\epsilon_r \equiv \delta m_x/m_x = 0.3 \cdot10^{-6}$ is the relative standard deviation at the measurement of the electron and proton mass, then $\delta t \sim 1.6 \cdot 10^6~ \delta t_0$. In the case of the electron, $ \delta t_0 = \ell/c \sim  10^{-21}$ s is comparable to the estimates of the "jump time" $\tau_J \sim 10^{-20}$ s for atomic transitions \cite{schul}. These change however the electron wave function over a distance larger than $ 10^3 \ell$, so that $\delta t$ could be a reasonable upper limit for $\tau_J$. 
\section{Summary and conclusions}  
The phase-space description of the physical states provides the conceptual framework for nonrelativistic many-body theory, statistical mechanics and canonical quantization. The asymmetry between time and the usual phase-space coordinates requires though "the second quantization", to obtain a Lorentz-covariant quantum theory. \\ \indent 
For classical relativistic systems, we may also extended the usual phase-space by energy and time, as canonical variables. In this work, scalar probability waves, describing free particles, are associated with functional coherent states for the Liouville equation (\ref{leq0}) in the extended phase-space.  
\\ \indent
The canonical equations of motion for a relativistic particle are presented in Sect. 2. As the usual time becomes a coordinate, the trajectories are parameterized by a variable $u$ called universal time. Action waves $n^e(q^e,u)^{[S]}$, associated with specific coherent solutions (\ref{cs1}) of (\ref{leq0}), are discussed in Sect. 3. These solutions are localized in the momentum space, and propagate according to the continuity (\ref{cont}) and Hamilton-Jacobi (\ref{hj}) relativistic equations. It is shown that in a finite system the universal time is the expectation value of the time coordinate in the intrinsic frame. 
\\ \indent
The transition from $n^e(q^e,u)^{[S]}$ to the quantum waves $\Psi(q^e,u)$ is discussed in Sect. 4. Presuming the existence of minimum space and time intervals, the action distributions (\ref{f0}) take the form of the relativistic quantum distributions (\ref{fq}). In the quantum case, the coherent solution of the Liouville equation is defined by the extended Wigner transform of the wave function provided by the relativistic Schr\"odinger equation (\ref{rse1}). For an ideal non-interacting, "static" system, this reduces to the Klein-Gordon equation (\ref{kg}). Most physical situations are though nonstationary, as all mesons undergo irreversible decay, while in the nonrelativistic quantum theory time is the same as in classical mechanics. When time is described quasiclassically, by a coherent wave-packet, then over large intervals compared to the width, the extended formalism reduces to the usual nonrelativistic quantum dynamics. 
\section{Appendix 1: Galilei and Lorentz actions} 
Let us consider a particle with mass $m$, described by the Cartesian phase-space coordinates $({\bf q}, {\bf p})$. An infinitesimal Galilei transformation $\Gamma_Q : {\sf R}^3 \times {\sf R} \rightarrow {\sf R}^3 \times {\sf R}$, acting both on the coordinate space ($Q = {\sf R}^3$) and time (${\sf R}$), is defined by $ [ {\bf q'}, t']= [ {\bf q}, t ]+ \gamma (\xi, {\bf d}, {\bf v}, \tau) [ {\bf q},t ]$, where 
\begin{equation} 
\gamma (\xi, {\bf d}, {\bf v}, \tau) [ {\bf q} , t ] = [\xi {\bf q} - {\bf d} - t {\bf v}, - \tau]~~. \label{gal} 
\end{equation} 
The algebra $g$ of the Galilei group is isomorphic to $so(3) + {\sf R}^7$, and $\gamma \in g$ is specified by $ \xi \in so(3)$, ${\bf d} \in {\sf R}^3$, ${\bf v} \in {\sf R}^3$ and $\tau \in {\sf R}$. The parameters $\xi$, ${\bf d}$ and ${\bf v}$ correspond to static rotations, translations and boost, respectively, of the space coordinates, while $\tau$ describes translations along the time axis. \\ \indent 
The action $\Gamma_Q$ of the Galilei group can be lifted to an action $\Gamma_M$ on the phase-space $M=T^* {\sf R}^3$, by assuming that at the transformation specified by (\ref{gal}), the momentum also changes to 
\begin{equation}
{\bf p}'= {\bf p} + \xi {\bf p} - m {\bf  v}~~. \label{p'}
\end{equation}
However, as the boost transformations depend on time explicitly, and 
\begin{equation}
p'_0=p_0 + {\bf v} \cdot {\bf  p} /c ~~, \label{p0}
\end{equation}
($p_0 = - E /c$), $\Gamma_Q$ can be lifted directly to an action $\Gamma_{M^e}$ on
the extended phase-space $M^e$ of Sect. 2. If the coordinates on $M^e$ are represented as column vectors 
$$ 
\tilde{q} = \left[ \begin{array}{c} {\bf q} \\ q_0 \end{array} \right]~~,~~ \tilde{p} = \left[ \begin{array}{c} {\bf p} \\ p_0 \end{array} \right] 
$$ 
then the infinitesimal transformation $\Gamma_{M^e}$ defined by (\ref{gal}), (\ref{p'}) and (\ref{p0}) takes the form of a canonical transformation
\begin{equation} 
\left[ \begin{array}{c} \tilde{q}' \\ \tilde{p}' \end{array} \right] = \left[ \begin{array}{c} \tilde{q} \\ \tilde{p} \end{array} \right] + \left[ \begin{array}{c} - \tilde{Y} \\ - \tilde{X} \end{array} \right] + \left[ \begin{array}{cc} - \hat{a}^T & \hat{c} \\ - \hat{b} & \hat{a} \end{array} \right] 
\left[ \begin{array}{c} \tilde{q} \\ \tilde{p} 
\end{array} \right]~~, \label{ict}
\end{equation} 
with 
\begin{equation} 
\tilde{X} = \left[ \begin{array}{c} m {\bf v} \\ 0 \end{array} \right] ~,~ \tilde{Y} = \left[ \begin{array}{c} {\bf d} \\ \tau \end{array} \right] ~,~ \hat{a} = \left[ \begin{array}{cc} \xi & {\bf 0} \\ {\bf v}/c & 0 \end{array} \right]~, \label{xya}
\end{equation}
and $\hat{b}=\hat{c}$, $4 \times 4$ zero matrices. \\ \indent
The mass $m$, introduced with the lift (\ref{p'}) is the positive, isotropic, inertial parameter for a Hamiltonian $H$ defined on $M$. In the case of a Hamiltonian $H^e$ defined on $M^e$, there is also an inertial parameter specified by the dependence of $H^e$ on the additional momentum component ($p_0$). Following \cite{et}, this new inertial parameter is taken for simplicity as  $ \pm m$, with $+$ or $-$ sign in the isotropic, respectively quasi-isotropic case. This yields a relationship of the form $p_0 = \pm mc$, or $ E= \mp m c^2$, which shows that the lift (\ref{ict}) of $\Gamma_Q$ should be obtained by placing the velocity ${\bf v}$ from (\ref{p'}) in the matrix $\hat{a}$, instead of $\tilde{X}$. In this case, (\ref{xya}) is replaced by
\begin{equation} 
\tilde{X} = \left[ \begin{array}{c} 0 \\ 0 \end{array} \right] ~~,~~ \tilde{Y} = \left[ \begin{array}{c} {\bf d} \\ \tau \end{array} \right] ~~,~~   
\hat{a} = \left[ \begin{array}{cc} \xi &  \mp {\bf v} /c \\ {\bf v}/c & 0 \end{array} \right]~~. \label{loa}
\end{equation}
According to (\ref{ict}), the new element in $\hat{a}$ changes the Galilei action (\ref{gal}) of the inertial equivalence group, and in the case $E > 0$ it provides the Lorentz action  
\begin{equation} 
\gamma_L (\xi, {\bf d}, {\bf v}, \tau) [ {\bf q} , t ] = [\xi {\bf q} - {\bf d} - t {\bf v}, -  {\bf v} \cdot {\bf q} /c^2 - \tau] ~~. \label{lga}
\end{equation} 
To find the action of the Lorentz group, (\ref{lga}) should be integrated to finite transformations. Let us presume that $\xi =0$, ${\bf d} =0$, $\tau =0$, 
and decompose the vectors ${\bf q}$, ${\bf p}$ with respect to the versor ${\bf n}$ of the boost velocity as ${\bf q}={\bf q}_\perp + q_\parallel {\bf n} $, ${\bf p}= {\bf p}_\perp + p_\parallel {\bf n} $.
In the representation
$$ 
\tilde{q} = \left[ \begin{array}{c} {\bf q}_\perp \\ q_\parallel \\ q_0 \end{array} \right]~~,~~ \tilde{p} = \left[ \begin{array}{c} {\bf p}_\perp \\ p_\parallel \\ p_0 \end{array} \right] ~~,
$$ 
we get $\hat{a}= \rho \hat{a}_0 $, where $\rho \equiv \vert {\bf v} \vert /c$, 
\begin{equation} 
\hat{a}_0 = \left[ \begin{array}{cc} \hat{0}_\perp & \hat{0} \\ \hat{0} &  \hat{\sigma}_x \end{array} \right]~~, 
\end{equation}
$\hat{0}_\perp= \hat{0}$ are $2 \times 2$ zero matrices, and  
\begin{equation} 
\hat{\sigma}_x = \left[ \begin{array}{cc} 0 & 1 \\ 1 & 0  \end{array} \right]~~.  
\end{equation}
Because
$$
e^{\rho \hat{\sigma}_x } = \cosh \rho ~\hat{1} + \sinh \rho ~ \hat{\sigma}_x ~~, 
$$
then for a boost transformation with finite velocity  ${\bf V} = V {\bf n}$, the equations 
\begin{equation}
\frac{ d \tilde{q}}{d \rho} = - \hat{a}^T_0 \tilde{q} ~~,~~\frac{ d \tilde{p}}{d \rho} =  \hat{a}_0 \tilde{p}
\end{equation}  
can be integrated to ${\bf q}'_\perp = {\bf q}_\perp$, ${\bf p}'_\perp = {\bf p}_\perp$, and
$$
q'_\parallel = \cosh \rho ~q_\parallel - \sinh \rho~ q_0 ~~,~~q'^0 = \cosh \rho ~q_0 - \sinh \rho~ q_\parallel 
$$ 
$$
p'_\parallel = \cosh \rho ~p_\parallel + \sinh \rho ~p_0~~,~~p'_0 = \cosh \rho~ p_0 + \sinh \rho~ p_\parallel ~~.
$$
These expressions show clearly the invariance of the Poisson bracket in the extended phase-space, because if $\{q_\mu, q_\nu \}^e = \{p_\mu, p_\nu \}^e=0$, $\{q_\mu, p_\nu \}^e = \delta_{\mu \nu}$, then also $\{q_\mu', q_\nu' \}^e = \{p_\mu', p_\nu' \}^e=0$, $\{q_\mu', p_\nu' \}^e = \delta_{\mu \nu}$, $\mu. \nu=0,1,2,3$. \\ \indent
The parameter $\rho$ is related to the finite boost velocity $V$ by physical considerations, such as $V=c d q_\parallel / d q_0$ when $dq'_\parallel=0$.
The result $V= c \tanh \rho$ provides the standard Lorentz transformations  
\begin{equation} 
{\bf q}'_\parallel =  \frac{ {\bf q}_\parallel -  {\bf V} t}{ \sqrt{1 - V^2/c^2}  }~~,~~ t' = \frac{ t - {\bf V} \cdot {\bf  q}/c^2 }{ \sqrt{1 - V^2/c^2}} ~~, 
\end{equation} 
\begin{equation} 
{\bf p}'_\parallel =  \frac{ {\bf p}_\parallel - {\bf V} E/c^2}{ \sqrt{1 - V^2/c^2}  }~~,~~ E' = \frac{ E - {\bf V} \cdot {\bf  p} }{ \sqrt{1 - V^2/c^2}} ~~. 
\end{equation} 
For states with negative energy ($E<0$), $ \mp {\bf v}$ in (\ref{loa}) takes the $-$ sign, the hyperbolic functions become trigonometric functions, and the Lorentz group SO(3,1) is replaced by the rotation group in space-time SO(4) \cite{et}, isomorphic to SU(2)$\times$SU(2). 
\section{Appendix 2: The relativistic perfect gas}
For a nondegenerate gas of fermions with energy $\epsilon_p = \sqrt{{\bf p}^2 c^2 +m_0^2c^4}$ at equilibrium, the usual distribution function has the form \cite{somm} 
\begin{equation}
f_{\mu, T} ({\bf p}) = \frac{2}{h^3} e^{(\mu- \epsilon_p)/k_BT }~~, \label{fcl} 
\end{equation}
so that if ${\cal V}$ denotes the confinement volume, then
\begin{equation}
N = {\cal V} \int d^3 p~  f_{\mu, T} ({\bf p}) ~~,~~E = {\cal V} \int d^3 p ~ \epsilon_p f_{\mu, T} ({\bf p})
\end{equation}
are the number of particles, and the total energy, respectively. The function (\ref{fcl}) is also a stationary solution of the classical Fokker-Planck equation
\begin{equation}
\partial_t f + \frac{1}{m} {\bf p} \cdot \nabla f = \gamma \nabla_p \cdot (  \frac{\bf p}{m} + k_B T \nabla_p ) f  ~~,
\end{equation}  
with $m=\epsilon_p /c^2$. The energy can be expressed in the form 
\begin{equation}
E = \frac{ 8 \pi {\cal V}}{h^3 c^3}  e^{\mu/k_BT}  \int_{m_0 c^2}^\infty d \epsilon ~ g_T (\epsilon)~~,
\end{equation}
where $g_T (\epsilon) = \epsilon^2 \sqrt{ \epsilon^2 -m_0^2c^4} \exp(- \epsilon /k_BT)$. 
Here the upper integration limit $\epsilon_M$ was presumed infinite, although in most physical situations particles with high enough energy can escape the system before thermalization. Moreover, the opening of pair creation reaction channels \cite{jps} at $\epsilon =3 m_0c^2,5m_0c^2,...$ also affects the distribution. Therefore, a reasonable limit of the energy range for a stationary distribution with a well-defined number of particles is $\epsilon_M \approx 3 m_0c^2$, which corresponds to the maximum of $g_T(\epsilon)$ at $T =m_0c^2/k_B$, when the old sound velocity formula $v_s = \sqrt{k_BT/m_0} $ yields $v_s=c$.  \\ \indent
By the simple limitation of the energy range, the inverse of (\ref{fk}) can be expressed in terms of a multiple Fourier series. Thus, a real function $f(X)$ defined on the finite domanin $[- \Delta, \Delta]$ can be represented in the form
\begin{equation}
f(X) = \frac{1}{2 \Delta} \sum_{n= - \infty}^\infty e^{ - i n \pi X / \Delta} \tilde{f}_n~~,
\end{equation}
where
\begin{equation}
\tilde{f}_n  = \int_{- \Delta}^\Delta d X ~ e^{i n \pi X / \Delta} f (X)~~.
\end{equation}
 
\end{document}